\documentclass{pasj00}

\begin{document}

\SetRunningHead{K.Hayasaki}{A new mechanism for massive binary black hole evolution}
\Received{2008/5/19}
\Accepted{2008/?/?}

\title{A new mechanism for massive binary black hole evolution}

\author{Kimitake \textsc{Hayasaki}}

\altaffiltext{}{Yukawa Institute for Theoretical Physics, Kyoto University, \\
Oiwake-cho, 
Kitashirakawa, Sakyo-ku, Kyoto 606-8502 }
\email{kimitake@yukawa.kyoto-u.ac.jp}

%

\KeyWords{black hole physics -- accretion, accretion disks 
-- binaries:general -- galaxies:nuclei}

\maketitle

\begin{abstract}
A massive binary black hole (BBH) is inevitably formed 
in a merged galactic nucleus before the black holes finally 
merge by emitting the gravitational radiation.
However, it is still unknown 
how the BBH evolves after its semi-major axis reached 
to the sub-parsec/parsec scale 
where the dynamical friction with the neighboring stars is no longer effective 
(the so-called the final parsec problem).
In this paper, we propose a new mechanism 
by which the massive BBH can naturally coalesce within a Hubble time.
We study the evolution of the BBH with triple disks 
which are composed of an accretion disk around each black hole 
and one circumbinary disk surrounding them.
While the circumbinary disk removes the orbital angular momentum of the BBH via 
the binary-disk resonant interaction, the mass transfer from the circumbinary disk to 
each black hole adds some fraction of its angular momentum 
to the orbital angular momentum of the BBH.
We find that there is a critical value of the mass-transfer rate 
where the extraction of the orbital angular momentum from the BBH is balanced 
with the addition of the orbital angular momentum to the BBH.
The semi-major axis of the BBH decays with time 
whereas the orbital eccentricity of the BBH grows with time,  
if the mass transfer rate is smaller than the critical one, and vice versa.
Its evolutionary timescale is characterized by the product of 
the viscous timescale of the circumbinary disk and 
the ratio of the total black hole mass to the mass of the circumbinary disk.
Since a minimum value of the critical mass-transfer rate 
is larger than the Eddington accretion rate of massive black holes
with masses in the $10^{6}M_{\odot}$ to $10^{9}M_{\odot}$ range 
as far as the evolutionary timescale is shorter than a Hubble time, 
it is promising that the critical mass-transfer rate is larger than the mass transfer rate.
Most of massive BBHs, therefore, enable to merge within a Hubble time by the
proposed mechanism, which helps to solve the final parsec problem.

\end{abstract}



\section{Introduction}
\label{sec:intro}
Massive black holes in galactic nuclei 
are considered to have co-evolved with their host galaxies \citep{fm00,geb00,mag98}.
Since galaxies are well-known to evolve through frequent mergers, 
this strongly suggests that black hole growth is 
mainly caused by black hole mergers and subsequent accretion of gas \citep{yutre02,dim05}. 
If so, a massive binary black hole (BBH) is inevitably formed 
before the black holes merge by emitting the gravitational radiation.
Even if there are transiently triple massive black holes in a galactic nucleus, 
the system finally settles down to the formation of the massive BBH 
by merging of two black hole or by ejecting one black hole from the system 
via a gravitational slingshot \citep{iwa06}.
Recent hydrodynamic simulations showed the rapid BBH formation within several $\rm{Gyrs}$ 
by the interaction between the black holes and 
the surrounding stars and gas in gas-rich galaxy merger \citep{may07}.

It is widely accepted that massive BBH 
mainly evolves via three stages \citep{bege80,yu02}.
Firstly, each black hole sinks independently 
towards the center of the common gravitational potential 
due to the dynamical friction with neighboring stars.
When the separation between two black holes
becomes less than $1\,\rm{pc}$ or so, 
an angular momentum loss by the dynamical friction 
slows down due to the depletion of the stars on orbits
intersecting the BBH \citep{sas74}.
This is the second evolutionary stage.
Finally, the BBH coalesces rapidly
if the semi-major axis decreases to the point where
the emission of the gravitational radiation becomes an efficient mechanism
to remove the orbital angular momentum of the BBH.
The transition from the second stage to the final stage is, however, 
considered to be the bottleneck of evolutionary path for the BBH to coalesce
because of cutting off the supply of the stars on intersecting orbits.
This is called the final parsec problem (see \cite{mm05} for a review).

Many authors have tackled the final parsec problem in the context of the interaction
between the black holes and the stars, but there has been still extensive discussions
\citep{roos81, makino97, q97, milo01, makino04, me07, se07}.
There is other possible way to extract the energy and 
the angular momentum from the BBH
by the interaction between the black holes and the gas surrounding them.
In some cases of circular binaries with extreme mass ratio,
the secondary black hole could be embedded 
in the gas disk around the primary black hole 
and then migrates to the primary black hole.
This kind of the binary-disk interaction  
could also be the candidate to resolve the final parsec problem \citep{iv99,go00,armi1,armi2}.

\begin{figure}
\resizebox{\hsize}{!}{
\includegraphics*[width=56mm]{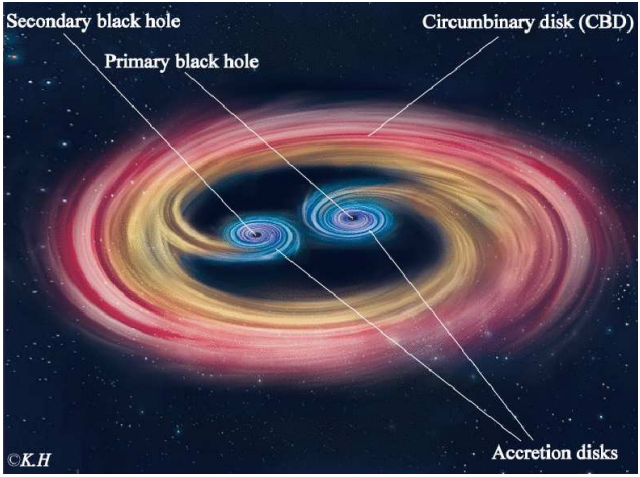}
}
  \caption{An artist's impression of a massive BBH with triple disks 
           on a parsec/subparsec scale of merged galactic nucleus.
           The BBH is surrounded by a circumbinary disk (CBD), from which
           the gas transfers to the central binary, and then 
           the accretion disk is formed around each black hole.
          }
  \label{fig:system}
\end{figure}

\citet{haya07} found that if the BBH is surrounded by the gaseous disk 
with the nearly Keplerian rotation (i.e. the circumbinary disk: CBD), 
the gas can be transferred from the CBD to each black hole.
The mass transfer leads to the formation of the accretion disk around each black hole \citep{haya08}, and then the BBH system finally has the triple disks
which are composed of the accretion disk around each black hole and one circumbinary disk surrounding them 
(see Fig.~\ref{fig:system} for a schematic view of the BBH system).
There is, however, little known how the BBH evolves in such a triple-disk system.
We, therefore, study the evolution of the massive BBH interacting with the triple disks.
Although we mainly discuss the case of a massive BBH system, our results can be applied to other possible context 
such as the compact binaries, the young binary star formation, and 
the extrasolar planet formation because of the scaling nature.
The plan of this paper is organized as follows. 
In Section~2, we describe the derivation of basic equations governing the orbital evolution of the BBH
 and the formulation of the gravitational torque of the BBH which acts on the CBD. 
The solutions for their evolutionary equations are, then, reported in Section~3.
Section~4 is devoted to discussions, and finally we summarize results in Section~5.

\section{Basic equations}

We assume that 
two black hole are gravitationally bounded and their motions are 
followed by the Kepler's third law, by which the BBH is defined.
The total energy of the BBH $E_{\rm{b}}$ is written by
\begin{equation}
E_{\rm{b}} = -\frac{GM_{\rm{bh}}\mu}{2a},
\label{eq:energy}
\end{equation}
where $a$ is the semi-major axis of the BBH and 
$\mu=M_{\rm{1}}M_{\rm{2}}/M_{\rm{bh}}$ is the reduced mass of the BBH.
Here, $M_{\rm{bh}}=M_{\rm{1}}+M_{\rm{2}}$ is total black hole mass, 
$M_{\rm{1}}$ is the primary black hole mass, 
and $M_{\rm{2}}$ is the secondary black hole mass.
By differentiating both sides of above equation, 
the energy dissipation rate of the orbital motion can be obtained as
\begin{equation}
\frac{\dot{E_{\rm{b}}}}{E_{\rm{b}}} = -\frac{\dot{a}}{a} + \frac{\dot{M_{1}}}{M_{1}} + \frac{\dot{M_{2}}}{M_{2}},
\label{eq:ebdot}
\end{equation}
where $\dot{M_{1}}$ is the mass accretion rate of the primary black hole, $\dot{M_{2}}$ is the mass accretion rate of the secondary black hole.
The dot over each physical quantity shows the time differentiation, unless otherwise noted.

The orbital angular momentum of the BBH $J_{\rm{b}}$ is written by 
\begin{eqnarray}
J_{\rm{b}}
=\mu{a^2}\Omega_{\rm{b}}\sqrt{1-e^2},
\label{eq:jb}
\end{eqnarray}
where $e$ is the orbital eccentricity,  
$\Omega_{\rm{b}}$ is the angular frequency of the BBH.
The relationship between the orbital energy $E_{\rm{b}}$ and the orbital angular momentum $J_{\rm{b}}$ 
can be written, from equation (\ref{eq:energy}) and (\ref{eq:jb}), 
\begin{eqnarray}
E_{\rm{b}}=-\frac{\Omega_{\rm{b}}J_{\rm{b}}}{2\sqrt{1-e^2}}.
\label{eq:ebjb}
\end{eqnarray}
The change rate of the orbital angular momentum $\dot{J}_{\rm{b}}$ can be written as
\begin{eqnarray}
\frac{\dot{J}_{\rm{b}}}{J_{\rm{b}}}=  \frac{\dot{a}}{2a} -\frac{e\dot{e}}{(1-e^2)}
-\frac{\dot{M}_{\rm{bh}}}{2M_{\rm{bh}}} +
\frac{ \dot{ M}_{1}}{M_{1}} + \frac{\dot{M}_{2}}{M_{2}},
\label{eq:jbdot}
\end{eqnarray}
where $\dot{M}_{\rm{bh}}=\dot{M}_{1}+\dot{M}_{2}$ is the total mass-accretion rate. 
The orbital evolution of the BBH is generally  governed by equations (\ref{eq:energy})-(\ref{eq:jbdot}).
Since the total mass accretion rate  is much lower than the black hole mass, 
the second term and the third term in the right hand-side 
of equations (\ref{eq:ebdot}) and (\ref{eq:jbdot}) can be
neglected. Actually, \citet{haya08} showed by the Smoothed Particle Hydrodynamics simulations 
that the mass accretion rate is two orders of magnitude lower than the mass transfer rate.
Therefore, a set of the basic equations of the orbital evolution is finally given by
\begin{eqnarray}
\frac{\dot{a}}{a}=-\frac{\dot{E_{\rm{b}}}}{E_{\rm{b}}}
\label{eq:oevo}
\end{eqnarray}
and
\begin{eqnarray}
\frac{e\dot{e}}{1-e^2}=-\frac{\dot{J}_{\rm{b}}}{J_{\rm{b}}}-\frac{\dot{E}_{\rm{b}}}{2E_{\rm{b}}},
\label{eq:edot}
\end{eqnarray}
respectively.


\subsection{Interaction between the disk and the binary black hole}

We consider the resonant interaction between the BBH and  the CBD.
The CBD is assumed to be geometrically thin, be aligned with the orbital plane of the BBH, 
and have nearly Keplerian rotation with no self-gravitation.
The binary potential can be expanded by Fourier double series,
\begin{equation}
\Phi(r,\theta,t) = \sum_{m,l}\phi_{ml}(r)\exp[i(m\theta-l\Omega_{\rm{b}}t)],
\label{eq:bp}
\end{equation}
where $m$ is azimuthal number, $l$ is the time-harmonic number.
The potential component $\phi_{m,l}$, can be written by
\begin{equation}
\phi_{ml}(r)=\frac{1}{2\pi^2}\int_{0}^{2\pi}d(\Omega_{\rm{b}}t)\int_{0}^{2\pi}\Phi\cos(m\theta-l\Omega_{\rm{b}}t),
\label{eq:bp1}
\end{equation}
which give rise to a corotational resonance at radius where { $\Omega_{\rm{p}}=\Omega$}
and the Lindblad resonances at radii $\Omega_{\rm{p}}=\Omega\pm(\kappa/m)$.
{ Here, $\Omega_{\rm{p}}$ is the pattern frequency of the binary potential which is defined by}
\begin{equation} 
\Omega_{\rm{p}}=\frac{l}{m}\Omega_{\rm{b}},
\label{eq:pattern}
\end{equation}
$\kappa$ is the epicyclic frequency, and 
the upper and lower signs correspond to the outer Lindblad
resonance (OLR) and the inner Lindblad resonance (ILR), respectively.
The radii of these resonances for the CBD are given by
\begin{eqnarray}
r_{\rm{CR}} = \left(\frac{m}{l}\right)^{2/3}a
\label{eq:co}
\end{eqnarray}
and
\begin{eqnarray}
r_{\rm{LRs}}=\left(\frac{m\pm1}{l}\right)^{2/3}a.
\label{eq:lrs}
\end{eqnarray}
Here, the epicyclic frequency $\kappa\sim\Omega$ for the nearly Keplerian disks. 
The standard formula for toques at the LR is given by \citet{gt79}
\begin{equation}
T_{ml}^{\rm{LRs}}=\frac{m(m\pm1)\pi^2\Sigma(\lambda-2m)^2\phi_{ml}^2}
{3l^2\Omega_{\rm{b}}^2},
\label{eq:sfgt79}
\end{equation}
where $\lambda=d\ln\phi_{ml}/d\ln{r}|_{\rm{LRs}}$. If $m>1$, 
$\lambda=-(m+1)$ for the OLR in the CBD (cf. \citet{al94}).
Similarly, the torque at the CR is written as 
\begin{equation}
T_{ml}^{\rm{CR}}=\frac{2m^3\pi^2\Sigma\phi_{ml}^2}
{3l^2\Omega_{\rm{b}}^2}.
\label{eq:sf2gt79}
\end{equation}
Note that the angular momentum is added from the binary to the CBD
{ via the LRs and the CR}.
On the other hand, the viscous torque formula derived by \citet{lp86} is written as
\begin{equation}
T_{\rm{vis}}=3\pi\alpha_{\rm{SS}}\Sigma\Omega^2r^4\left(\frac{H}{r}\right)^2,
\end{equation}
where $\alpha_{\rm{SS}}$ is the Shakura-Sunyaev viscosity parameter \citep{ss73}.
Since the viscous torque removes the angular momentum from the CBD,
the CBD is truncated and forms the gap between the CBD and the binary 
if the viscous torque is less than the resonant toque.
The inner edge of the CBD is formed at the given resonance radius
where the viscous torque is balanced with the resonant torque: 
\begin{equation}
T_{\rm{vis}} =\sum_{ml}T_{ml}^{\rm{OLR}}+\sum_{ml}T_{ml}^{\rm{ILR}}+\sum_{ml}T_{ml}^{\rm{CR}}\simeq\sum_{ml}T_{ml}^{\rm{OLR}},
\label{eq:crit}
\end{equation}
where the summation is taken over all combination of 
(m, l) which give the same resonance radius. 
Actually, criterion (\ref{eq:crit}) is determined only by the 
viscous torque and the torques from the OLR of the lowest-order potential component, 
because the torques from the OLR dominate those from the CR in the CBD 
and high-order potential components contribute little to the total torque, 
even if the eccentricity of the binary is not small \citep{gt79,al91,al94}. 
Following \citet{al94}, the inner edge of the CBD $r_{\rm{in}}$ 
is approximately determined by $(m,l)=(2,1)$ resonant torque 
in the binary with a low eccentricity: $r_{\rm{in}}=(m+1/l)^{2/3}a\simeq2.08a$.
It is confirmed by \citet{haya07} that this value is also applicable 
for the BBH with the moderate eccentricity $e=0.5$.
Therefore, the BBH transfers most angular momentum to the CBD 
via the 1:3 resonant radius $r_{\rm{in}}\simeq2.08a$.
{ Below,} the set of $(m,l)$ is regarded as $(2,1)$, unless otherwise noted.


\section{Orbital evolution of massive binary black holes}

The long-term evolution of the semi-major axis and of the orbital eccentricity 
are mainly driven by the resonant interaction between the BBH and
the CBD. In this section, we firstly describe the evolutionary 
relation between the semi-major axis 
and the orbital eccentricity, the evolution of the semi-major axis, 
the evolution of the orbital eccentricity,  and finally the effect of the mass transfer
from the CBD on the orbital evolution of the BBH.

\subsection{Relation between the semi-major axis and the orbital eccentricity}

\begin{figure}
\resizebox{\hsize}{!}{
\includegraphics*[width=86mm]{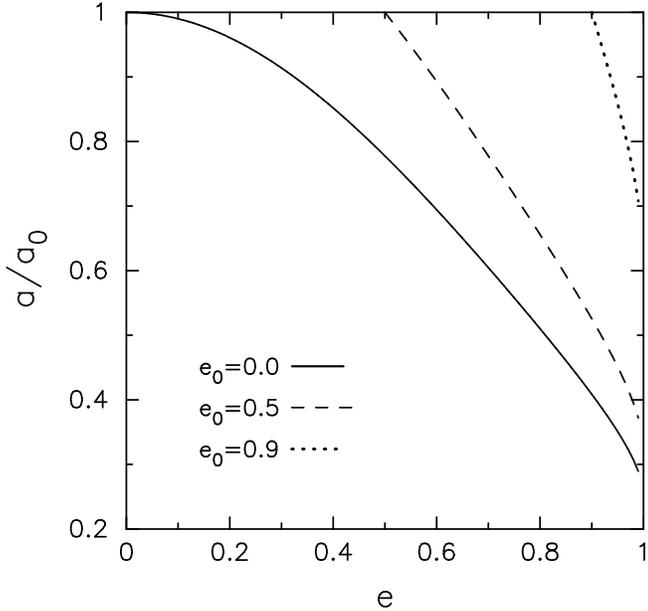}
}
\caption{
Evolution of the semi-major axis $a$ as a function of the orbital eccentricity $e$.
The semi-major axis, $a$, is normalized by the initial value of the semi-major axis $a_{0}$. 
The initial value of the orbital eccentricity is indicated by $e_{0}$.
The solid line, the dashed line, and the dotted line are the $a$-$e$ relations 
of $e_{0}=0.0$, of $e_{0}=0.5$, and of $e_{0}=0.9$, respectively. 
}
\label{fig:ae}
\end{figure}

As the motion of the BBH slowly vary with time by resonantly interacting with the CBD,
there is an adiabatic invariant which can be regards as the 
orbital angular momentum of the BBH (\cite{ll60}). 
In such the system, the energy dissipation rate of the BBH
is proportional to the product of the change rate of the adiabatic invariant
and { the characteristic frequency of the system, e.g., $\Omega_{\rm{p}}$}(cf. \cite{lubow96}):
\begin{eqnarray}
\dot{E_{\rm{b}}}=\Omega_{\rm{p}}\dot{J}_{\rm{b}}.
\label{eq:edr}
\end{eqnarray}
From equations (\ref{eq:ebjb}), (\ref{eq:pattern}), and (\ref{eq:edr}), equation (\ref{eq:oevo}) is rewritten as 
\begin{eqnarray}
\frac{\dot{a}}{a}
=2\frac{l}{m}\frac{\dot{J}_{\rm{b}}}{J_{\rm{b}}}\sqrt{1-e^2},
\label{eq:oevo2}
\end{eqnarray}
which determines the evolution of the semi-major axis of the BBH.
Using  above equation, equation (\ref{eq:edot}) is rewritten as
\begin{eqnarray}
\frac{e\dot{e}}{1-e^2}
=-\left(1-\frac{l}{m}\sqrt{1-e^2}\right)\frac{\dot{J}_{\rm{b}}}{J_{\rm{b}}}.
\label{eq:edot2}
\end{eqnarray}
From equation (\ref{eq:oevo2}) and (\ref{eq:edot2}), we can obtain the differential equation relating $a$ to $e$:
\begin{eqnarray}
\frac{\dot{a}}{a}=
-2\frac{l}{m}\frac{e\dot{e}}{\sqrt{1-e^2}}\biggl/\left(1-\frac{l}{m}\sqrt{1-e^2}\right).
\label{eq:aedif}
\end{eqnarray}
The above equation can be integrated, and then, 
$a$ can be expressed as the function of $e$:
\begin{eqnarray}
\frac{a}{a_{0}}=\left(1-\frac{l}{m}\sqrt{1-e_{0}^2}\biggl/1-\frac{l}{m}\sqrt{1-e^2}\right)^2,
\label{eq:ae}
\end{eqnarray}
where $a_{0}$ and $e_{0}$ are the initial value of the semi-major axis and the orbital eccentricity,
respectively.
Figure~\ref{fig:ae} exhibits the evolution of the semi-major axis $a$ normalized by $a_{0}$ 
as a function of the orbital eccentricity $e$. 
It is noted from the figure that the semi-major axis decays with time 
as the orbital eccentricity grows with time, and vice versa.
When the BBH evolves towards a black hole merger,
the eccentricity growth is faster than the orbital decay 
because the orbital eccentricity is more rapidly close to $1.0$ than 
the semi-major axis is close to 1.0.

\subsection{Evolution of the semi-major axis}
\begin{figure}
\resizebox{\hsize}{!}{
\includegraphics*[width=106mm]{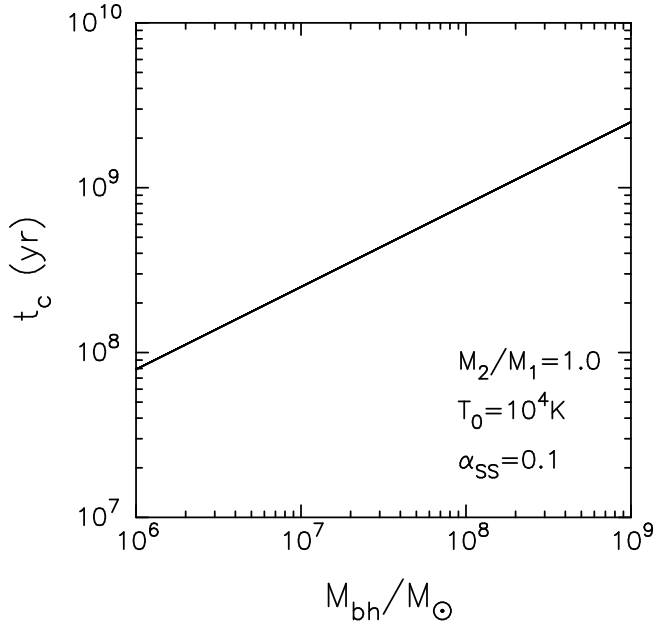}
}
\caption{
Characteristic timescale, $t_{\rm{c}}$, of massive BBH evolution
as a function of the total black hole mass in units of the solar mass, $M_{\rm{bh}}/M_{\odot}$, 
in the case of $M_{\rm{CBD}}/M_{\rm{bh}}=10^{-2}$, 
the Sakura-Sunyaev viscosity parameter $\alpha_{\rm{SS}}=0.1$, 
the black-hole mass ratio $M_{2}/M_{1}=1.0$, and the disk temperature $T_{0}=10^{4}K$.
}
\label{fig:tc}
\end{figure}

Assuming that the CBD completely absorbs the
orbital angular momentum of the BBH and is quasi-steady state,
the change rate of the orbital angular momentum $\dot{J}_{\rm{b}}$ can be written as $\dot{J}_{\rm{b}}=-\dot{J}_{\rm{CBD}}$, where $\dot{J}_{\rm{CBD}}$ is the change rate of the angular momentum of the CBD.
\begin{eqnarray}
\frac{\dot{a}}{a}=-2\frac{l}{m}\frac{\dot{J}_{\rm{CBD}}}{J_{\rm{b}}}\sqrt{1-e^2},
\label{eq:oevo3}
\end{eqnarray}
Since the orbital angular momentum is always transferred 
from the BBH to the CBD via the inner edge of the CBD $r_{\rm{in}}$, 
$\dot{J}_{\rm{CBD}}$ can be approximately estimated as
\begin{eqnarray}
\dot{J}_{\rm{CBD}}&\simeq&{T_{\rm{vis}}}|_{r=r_{\rm{in}}}=
3\left(\frac{m+1}{l}\right)^{1/3}\frac{(1+q)^2}{q} \nonumber \\
&\times&
\frac{M_{\rm{CBD}}}{M_{\rm{bh}}}\frac{1}{\sqrt{1-e^2}}\frac{J_{\rm{b}}}{\tau_{\rm{vis}}^{\rm{CBD}}},
\label{eq:jCBD}
\end{eqnarray}
where $q$ is the mass ratio of the secondary black hole to the primary black hole, {
the scope of which is $0.1\lesssim{q}\le1.0$ because the CBD is not truncated and thus the triple disk model is broken down if q is less than 0.1.}
Also, $M_{\rm{CBD}}\sim\pi r_{\rm{in}}^{2}\Sigma$, and $\tau_{\rm{vis}}^{\rm{CBD}}$ is the viscous timescale of the CBD measured at $r=r_{\rm{in}}$:
\begin{eqnarray}
\tau_{\rm{vis}}^{\rm{CBD}}&\sim&4.8\times10^{3}
\left(\frac{m+1}{l}\right)^{1/3}\left(\frac{0.1}{\alpha_{\rm{SS}}}\right) \nonumber \\
&\times&
\left(\frac{10^{4}\rm{K}}{T_{\rm{in}}}\right)\left(\frac{M_{\rm{bh}}}{M_{\odot}}\right)^{1/2}\left(\frac{a}{a_{0}}\right)^{1/2}\hspace{2mm}[\rm{yr}], 
\end{eqnarray}
where the initial value of a semi-major axis $a_{0}=1\rm{pc}$, 
and the CBD is assumed to be isothermal for simplicity.
Equation (\ref{eq:oevo3}) is finally written as
\begin{equation}
\frac{\dot{a}}{a}
=-\frac{2}{t_{\rm{c}}}\left(\frac{a}{a_{0}}\right)^{-1/2}.
\label{eq:adot}
\end{equation}
Here $t_{\rm{c}}$ shows a characteristic timescale of the evolution of the BBH, which is defined by
\begin{equation}
t_{\rm{c}}
=\frac{\tau_{\rm{vis},0}^{\rm{CBD}}}{{\zeta}}
\sim
\left(\frac{M_{\rm{bh}}}{M_{\odot}}\right)^{1/2}\left(\frac{M_{\rm{CBD}}}{M_{\rm{bh}}}\right)^{-1}
\label{eq:tc}
\end{equation}
where $\tau_{\rm{vis},0}^{\rm{CBD}}$ is the viscous timescale of the CBD when $a=a_{0}$, and 
${\zeta}$ is independent of both the semi-major axis $a$ and time $t$:
\begin{eqnarray}
{ \zeta}=3\frac{l}{m}\left(\frac{m+1}{l}\right)^{1/3}\frac{(1+q)^2}{q}\frac{M_{\rm{CBD}}}{M_{\rm{bh}}}
\sim\frac{M_{\rm{CBD}}}{M_{\rm{bh}}}.
\label{eq:lambda}
\end{eqnarray}
Figure~\ref{fig:tc} represents the characteristic timescale $t_{\rm{c}}$ of the massive BBH evolution
as function of the black hole mass in unit of the solar mass. 
Here, we set the $T_{0}=10^{4}K$ and $\alpha_{\rm{SS}}=0.1$, 
which are typical values for a disk in active galactic nuclei (AGNs).
The ratio of the CBD mass  to the black hole mass is set as 
$M_{\rm{CBD}}/M_{\rm{bh}}=10^{-2}$, which ensures that the CBD is stable for the self-gravitation 
because Toomre's $Q$ value is $Q=(M_{\rm{CBD}}/M_{\rm{bh}})^{-1}(H/r)>1$ \citep{toomre64}. 
The solid line denotes $t_{\rm{c}}$ in the range $M_{\rm{bh}}=10^{6}$-$10^{9}M_{\odot}$. 
As seen in Fig.~\ref{fig:tc}, $t_{\rm{c}}$ is the longer as the black hole is more massive.

\begin{figure*}
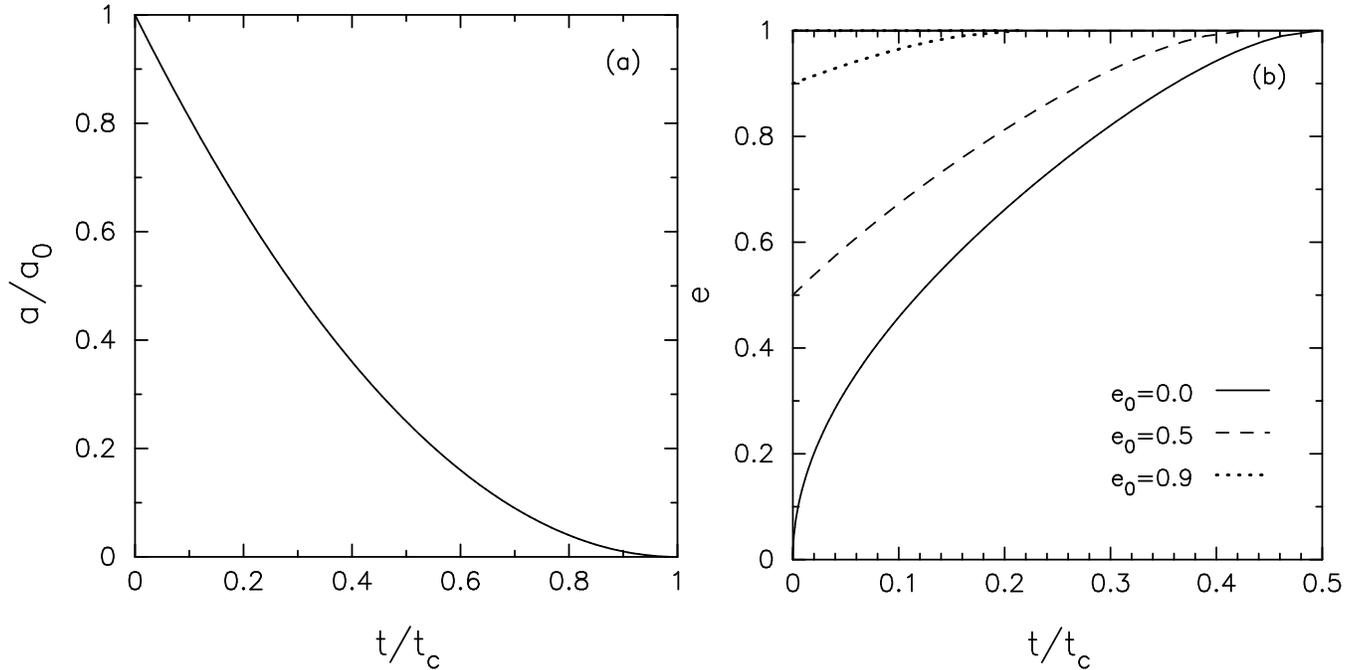

\resizebox{\hsize}{!}{
\includegraphics*[width=89mm]{kh4a.eps}
\includegraphics*[width=86mm]{kh4b.eps}
}
\caption{
({\it a}) Evolution of a semi-major axis normalized 
by the initial value of the semi-major axis, $a_{0}$, 
of the BBH for $0\le{t}\le{t_{\rm{c}}}$.
The solid line shows the evolution of the semi-major axis. 
({\it b}) Evolution of an orbital eccentricity of the BBH for $0\le{t}\le{t_{\rm{c}}}$.
The solid thin line, the dashed line and the dotted line show the evolution of 
the orbital eccentricity in the case of an initial value 
of the orbital eccentricity $e_{0}=0.0$, $e_{0}=0.5$ and $e_{0}=0.9$, respectively.
}
\label{fig:atet}
\end{figure*}


Integrating equation (\ref{eq:adot}),
\begin{equation}
\frac{a}{a_{0}}=\left(1-\frac{t}{t_{\rm{c}}}\right)^{2},
\label{eq:aevo}
\end{equation}
which determines the evolution of the semi-major axis of BBH.
Figure~\ref{fig:atet}{\it a} shows the evolution of the semi-major axis $a$ 
normalized by the initial value of the semi-major axis $a_{0}$. 
It is noted from the figure that the semi-major axis rapidly decays with time 
regardless of the orbital eccentricity, and the BBH finally merges at $t=t_{\rm{c}}$.


\subsection{Evolution of the orbital eccentricity}
\label{sec:et}
Substituting equation (\ref{eq:adot}) into equation (\ref{eq:aedif}) can be written as
\begin{eqnarray}
\frac{e\dot{e}}{\sqrt{1-e^2}}\biggl/\left(1-\frac{l}{m}\sqrt{1-e^2}\right)=\frac{m}{l}\frac{1}{(t_{\rm{c}}-t)},
\label{eq:edot3}
\end{eqnarray}
which determines the evolution of the orbital eccentricity.
Integrating both sides, we can obtain the evolutionary timescale of the orbital eccentricity
\begin{eqnarray}
\frac{t}{t_{\rm{c}}}
&=&1-\left[
\left(1-\frac{l}{m}\sqrt{1-e_{0}^{2}}\right) 
\biggl{/}\left(1-\frac{l}{m}\sqrt{1-e^2}\right)\right]
\end{eqnarray}
The evolution of the orbital eccentricity $e$ is 
\begin{eqnarray}
e=\sqrt{(1+\eta(t))(1-\eta(t))},
\label{eq:et}
\end{eqnarray}
where $\eta$ is
\begin{eqnarray}
\eta(t)=\frac{m}{l}\left[1-\left(1-\frac{l}{m}\sqrt{1-e_{0}^2}\right)\left[\frac{t_{\rm{c}}}{t_{\rm{c}}-t}\right]\right].
\label{eq:eta}
\end{eqnarray}
Figure~\ref{fig:atet}{\it b} represents the evolution of the orbital eccentricity of the BBH
in the case of $e_{0}=0.0$, $e=0.5$, and $e=0.9$, respectively.
This figure clearly shows that the orbital eccentricity grows with time and is close to $1.0$ 
within $t=t_{\rm{c}}$ regardless of the initial value of the orbital eccentricity.
The higher value of the initial eccentricity is more rapidly close to 1.0.

\begin{figure*}
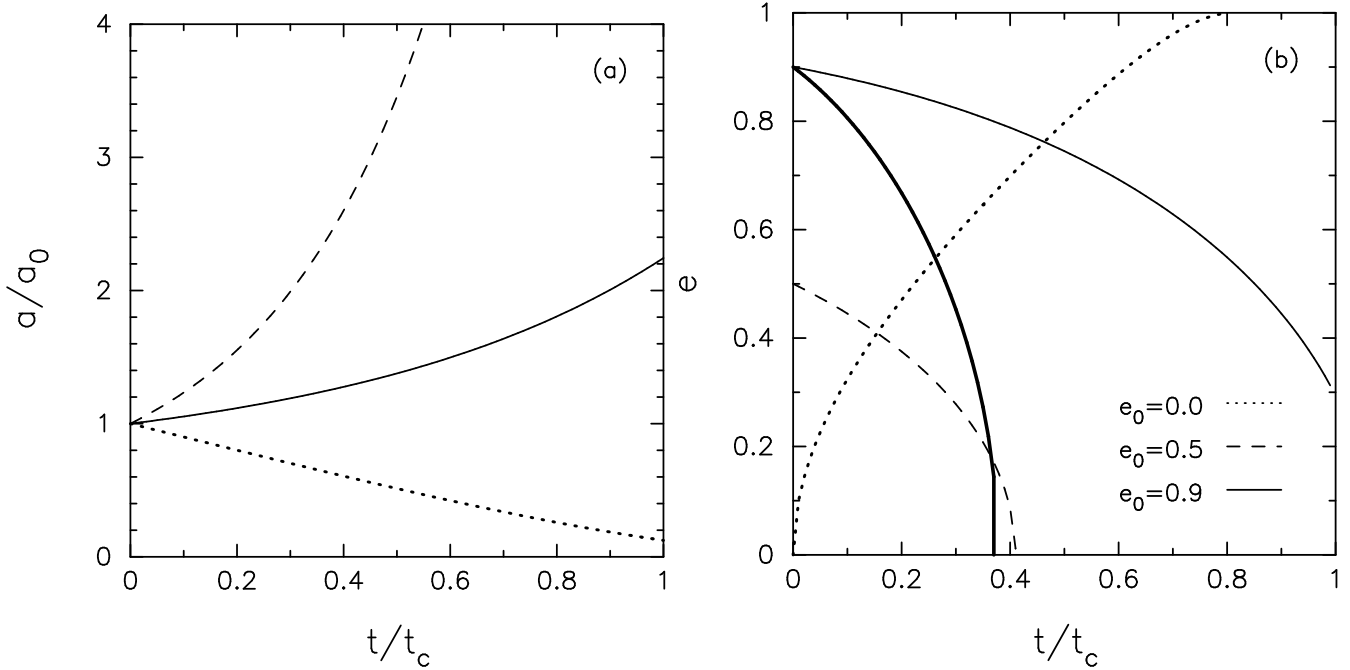

\resizebox{\hsize}{!}{
\includegraphics*[width=86mm]{kh5a.eps}
\includegraphics*[width=86mm]{kh5b.eps}
}
\caption{
({\it a}) Same format as Fig.~4{\it a}. 
The solid line, the dashed line, and the dotted line show the evolution of semi-major axis 
when $\langle\dot{M}_{\rm{T}}\rangle/{\dot{M}_{\rm{crit}}}=1.25$, 
$\langle\dot{M}_{\rm{T}}\rangle/{\dot{M}_{\rm{crit}}}=2.0$, and $\langle\dot{M}_{\rm{T}}\rangle/{\dot{M}_{\rm{crit}}}=0.5$, respectively.
({\it b}) Same format as Fig.~4{\it b}.
The solid thin line, the solid thick line, the dashed line, and the dotted line show the evolution
of the orbital eccentricity when 
$(e_{0},\langle\dot{M}_{\rm{T}}\rangle/{\dot{M}_{\rm{crit}}})=(0.9,1.25)$, 
$(e_{0},\langle\dot{M}_{\rm{T}}\rangle/{\dot{M}_{\rm{crit}}})=(0.9,2.0)$,  
$(e_{0},\langle\dot{M}_{\rm{T}}\rangle/{\dot{M}_{\rm{crit}}})=(0.5,1.25)$, and 
$(e_{0},\langle\dot{M}_{\rm{T}}\rangle/{\dot{M}_{\rm{crit}}})=(0.0,0.5)$, respectively.
}
\label{fig:atetmt}
\end{figure*}

\subsection{Effect of the mass transfer from the circumbinary disk}

\citet{haya07} found that
the mass transfer from the CBD to each black hole occurs every binary orbit.
Since the transfered mass has the relative angular momentum to each black hole,  
the accretion disk is formed around each black hole \citep{haya08}.
In this section, we investigate the effect of the mass transfer on 
the evolution of the massive BBH.

The averaged torque, which is added to the accretion disks by the mass transfer,
during one orbital period is given by 
\begin{eqnarray}
\dot{J}_{\rm{T}}
&\simeq&
\langle\dot{M}_{\rm{T}}\rangle{r_{\rm{in}}^{2}}\Omega_{\rm{in}} \nonumber \\
&=&\left(\frac{m+1}{l}\right)^{1/3}\frac{(1+q)^{2}}{q}\frac{\langle\dot{M}_{\rm{T}}\rangle}{M_{\rm{bh}}}\frac{J_{\rm{b}}}{\sqrt{1-e^2}},
\label{eq:jmt}
\end{eqnarray}
where $\langle\dot{M}_{\rm{T}}\rangle$ denotes the averaged mass-transfer rate during one orbital period,
which can be expressed as the sum of 
the averaged mass-transfer rate to the primary black hole (the primary transfer rate, 
$\langle\dot{M}_{\rm{T},1}\rangle$) and the averaged mass-transfer rate to the secondary black hole 
(the secondary transfer rate, $\langle\dot{M}_{\rm{T},2}\rangle$): $\langle\dot{M}_{\rm{T}}\rangle$=$\langle\dot{M}_{\rm{T},1}\rangle + \langle\dot{M}_{\rm{T},2}\rangle$. Unless otherwise noted, we call $\langle\dot{M}_{\rm{T}}\rangle$ the mass transfer rate.
The accumulated mass around each black hole accretes onto each black hole by which 
the angular momentum is transferred outward and is finally 
added to the binary due to the tidal torque (cf. \cite{kato08}). 
The torque added to the binary by this process, $\dot{J}_{\rm{add}}$, can be estimated as
\begin{eqnarray}
\dot{J}_{\rm{add}}=\dot{J}_{\rm{pbhd}} + \dot{J}_{\rm{sbhd}},
\label{eq:jadd}
\end{eqnarray}
where $\dot{J}_{\rm{pbhd}}$ and $\dot{J}_{\rm{sbhd}}$ are the torque added 
from the accretion disk around the primary black hole to the BBH (the primary torque) and 
the torque added from the accretion disk around the secondary black hole to the BBH (the secondary torque), 
respectively. By defining the ratio of the primary transfer rate to the secondary transfer rate, 
$q_{\rm{T}}=\langle\dot{M}_{\rm{T},2}\rangle/\langle\dot{M}_{\rm{T},1}\rangle$, 
the primary torque can be written by 
\begin{eqnarray}
\dot{J}_{\rm{pbhd}}
&=&
\langle\dot{M}_{\rm{T},1}\rangle\frac{P_{\rm{b}}}{\tau_{\rm{vis},1}}\sqrt{GM_{1}r_{\rm{c},1}}\left(1-\sqrt{\frac{r_{\rm{ms},1}}{r_{\rm{c}}}}\right)
\nonumber \\
&\simeq&
\frac{\langle\dot{M}_{\rm{T}}\rangle}{1+q_{\rm{T}}}\frac{P_{\rm{b}}}{\tau_{\rm{vis},1}}\sqrt{GM_{1}r_{\rm{c},1}},
\label{eq:jpbhd}
\end{eqnarray}
where $P_{\rm{b}}$ is the orbital period of the BBH, $\tau_{\rm{vis},1}$ is the viscous timescale measured at the circularization radius $r_{\rm{c},1}$ of the accretion disk around the primary black hole, and $r_{\rm{ms},1}$ is the radius of the marginally stable circular orbit of the primary black hole. 
Here, we approximate $\dot{J}_{\rm{pbhd}}$ by that  $r_{\rm{ms},1}$ is much smaller than $r_{\rm{c},1}$.
Similarly, the secondary torque can be written by
\begin{eqnarray}
\dot{J}_{\rm{sbhd}}\simeq\frac{q_{\rm{T}}\langle\dot{M}_{\rm{T}}\rangle}{1+q_{\rm{T}}}\frac{P_{\rm{b}}}{\tau_{\rm{vis},2}}\sqrt{GM_{2}r_{\rm{c},2}},
\label{eq:jsbhd}
\end{eqnarray}
where $\tau_{\rm{vis},2}$ is the viscous timescale measured at the circularization radius $r_{\rm{c},2}$ 
of the accretion disk around the secondary black hole. 
Substituting equations~(\ref{eq:jmt}) and (\ref{eq:jsbhd})-(\ref{eq:jpbhd}) into equation~(\ref{eq:jadd}), we can obtain
\begin{eqnarray}
\dot{J}_{\rm{add}}=f\dot{J}_{\rm{T}}.
\label{eq:jadd2}
\end{eqnarray}
Here, $f$ is the parameter which is defined as
\begin{eqnarray}
f&=&\frac{2\pi\alpha_{\rm{SS}}}{1+q_{\rm{T}}}\left[1+q_{\rm{T}}\left(\frac{c_{\rm{s},2}}{c_{\rm{s},1}}\right)^{2}\right]\left(\frac{m+1}{l}\right)^{-1/3}\left(\frac{c_{\rm{s},1}}{v_{\rm{b}}}\right)^2 \nonumber \\
&\simeq&2\pi\alpha_{\rm{SS}}\left(\frac{m+1}{l}\right)^{-1/3}\left(\frac{c_{\rm{s},1}}{v_{\rm{b}}}\right)^2,
\label{eq:f}
\end{eqnarray}
where $v_{\rm{b}}$ is the orbital velocity of the BBH, $c_{\rm{s},1}$ and $c_{\rm{s},2}$ 
are the sound velocity of the primary disk and the sound velocity of the secondary disk, respectively.
For simplicity, we assume that $c_{\rm{s},1}{\approx}c_{\rm{s},2}$ and 
the ratio of $c_{\rm{s},1}$ to $v_{\rm{b}}$ is constant during the evolution of the BBH.
The parameter $f$ shows how much the torque of the mass transfer is converted to the torque of the BBH.   
When $f$ equals to $1.0$, the torque of the mass transfer is completely converted to the torque of the BBH. 
The value of $f$ is, however, substantially much smaller than $1.0$ because the orbital velocity is much faster than the sound velocity 
at the circularization radius in general.

The angular momentum balance in the BBH system can be expressed as
\begin{eqnarray}
\dot{J}_{\rm{b}}
&=&-\dot{J}_{\rm{CBD}} + \dot{J}_{\rm{add}} \nonumber \\
&=&-\frac{m}{l}\frac{1}{t_{\rm{c}}}\left(1-\frac{\langle{\dot{M}_{\rm{T}}}\rangle}{\dot{M}_{\rm{crit}}}\left(\frac{a}{a_{0}}\right)^{1/2}\right)\left(\frac{a}{a_{0}}\right)^{-1/2} \nonumber \\
&\times&
\frac{J_{\rm{b}}}{\sqrt{1-e^2}},
\label{eq:jbdototal}
\end{eqnarray}
and thus, from equation~(\ref{eq:adot}), the differential equation of the semi-major axis can be written as
\begin{eqnarray}
\frac{\dot{a}}{a}
=-\frac{2}{t_{\rm{c}}}\left(1-\frac{\langle{\dot{M}_{\rm{T}}}\rangle}{\dot{M}_{\rm{crit}}}\left(\frac{a}{a_{0}}\right)^{1/2}\right)\left(\frac{a}{a_{0}}\right)^{-1/2},
\label{eq:mtadot}
\end{eqnarray}
where $\dot{M}_{\rm{crit}}$ is the critical mass transfer rate which can be defined as
\begin{eqnarray}
\dot{M}_{\rm{crit}}&=&\frac{3M_{\rm{CBD}}}{{f}\tau_{\rm{vis},0}^{\rm{CBD}}}
\nonumber \\
&=&\frac{1}{f}\frac{q}{(1+q)^2}\frac{m}{l}\left(\frac{m+1}{l}\right)^{-1/3}
\frac{M_{\rm{bh}}}{t_{\rm{c}}}.
\label{eq:mcrit}
\end{eqnarray}
Here,  we use equation~(\ref{eq:tc}) and (\ref{eq:lambda}).
Integrating both sides of equation~(\ref{eq:mtadot}), we obtain
\begin{eqnarray}
\frac{a}{a_{0}}
&=&\left(\frac{\langle{\dot{M}_{\rm{T}}}\rangle}{\dot{M}_{\rm{crit}}}\right)^{-2}
\nonumber \\
&\times&
\left[1-\left(1-\frac{\langle{\dot{M}_{\rm{T}}}\rangle}{\dot{M}_{\rm{crit}}}\right)
\exp\left(\frac{t}{t_{\rm{c}}}\frac{\langle{\dot{M}_{\rm{T}}}\rangle}{\dot{M}_{\rm{crit}}}\right)\right]^{2}.
\end{eqnarray} 
Figure~\ref{fig:atetmt}{\it a} displays the evolution of the semi-major axis of a massive BBH with triple disks.
It is noted from the figure that the semi-major axis $a$ decays with time when $\dot{M}_{\rm{crit}}>\langle\dot{M}_{\rm{T}}\rangle$, 
while the semi-major axis increases with time when $\dot{M}_{\rm{crit}}<\langle\dot{M}_{\rm{T}}\rangle$.
In the limit of $\langle\dot{M}_{\rm{T}}\rangle/\dot{M}_{\rm{crit}}\ll1.0$, the dotted line corresponds to the solid line of
Fig.~\ref{fig:atet}{\it a}.
The growth rate of the semi-major axis is more steep when $\langle\dot{M}_{\rm{T}}\rangle/\dot{M}_{\rm{crit}}=2.0$
than when $\langle\dot{M}_{\rm{T}\rangle}/\dot{M}_{\rm{crit}}=1.25$.
This means that the semi-major axis more rapidly grows with time as accretion disks are more massive.

Similarly, the differential equation of the orbital eccentricity $e$ is modified by the mass transfer as follows:
\begin{eqnarray}
\frac{e\dot{e}}{\sqrt{1-e^2}}
&\biggl/&\left(1-\frac{l}{m}\sqrt{1-e^2}\right)
\nonumber \\
&=&\frac{m}{l}\frac{1}{t_{\rm{c}}}
\left[
\frac{\langle{\dot{M}_{\rm{T}}}\rangle}{\dot{M}_{\rm{crit}}}\left(1-\frac{\langle{\dot{M}_{\rm{T}}}\rangle}{\dot{M}_{\rm{crit}}}\right) 
\exp\left(\frac{\langle{\dot{M}_{\rm{T}}}\rangle}{\dot{M}_{\rm{crit}}}\frac{t}{t_{\rm{c}}}\right)
\right] \nonumber \\
&\biggl/&\left[1-\left(1-\frac{\langle{\dot{M}_{\rm{T}}}\rangle}{\dot{M}_{\rm{crit}}}\right)\exp\left(\frac{t}{t_{\rm{c}}}\frac{\langle{\dot{M}_{\rm{T}}}\rangle}{\dot{M}_{\rm{crit}}}\right)\right]
\label{eq:edot3}
\end{eqnarray}
By integrating both sides of above equation, we can obtain the 
evolutionary equation of the orbital eccentricity.
\begin{eqnarray}
e=\sqrt{(1+\eta(t))(1-\eta(t))},
\label{eq:mtet}
\end{eqnarray}
where $\eta$ is
\begin{eqnarray}
\eta(t)
&=&
\frac{m}{l}-
\frac{m}{l}
\left(1-\frac{l}{m}\sqrt{1-e_{0}^2}\right)
\left(\frac{\langle{\dot{M}_{\rm{T}}}\rangle}{\dot{M}_{\rm{crit}}}\right) \nonumber \\
&\times&
\left[1-\left(1-\frac{\langle{\dot{M}_{\rm{T}}}\rangle}{\dot{M}_{\rm{crit}}}\right)\exp\left(\frac{t}{t_{\rm{c}}}\frac{\langle{\dot{M}_{\rm{T}}}\rangle}{\dot{M}_{\rm{crit}}}\right)\right]^{-1}.
\label{eq:mteta}
\end{eqnarray}
Figure~\ref{fig:atetmt}{\it b} represents the evolution of the orbital eccentricity
of a massive BBH with triple disks.
The figure shows that  the orbital eccentricity decays with time 
when $\langle{\dot{M}_{\rm{T}}}\rangle/\dot{M}_{\rm{crit}}>1.0$, whereas
it grows with time when $\langle{\dot{M}_{\rm{T}}}\rangle/\dot{M}_{\rm{crit}}<1.0$.
This is independent of the initial value of the orbital eccentricity $e_{0}$. 
The rate of evolution is more rapid as $\langle{\dot{M}_{\rm{T}}}\rangle/\dot{M}_{\rm{crit}}$
is larger than $1.0$ when $\langle{\dot{M}_{\rm{T}}}\rangle/\dot{M}_{\rm{crit}}>1$ 
and as $\langle{\dot{M}_{\rm{T}}}\rangle/\dot{M}_{\rm{crit}}$ is smaller than $1.0$
when $\langle{\dot{M}_{\rm{T}}}\rangle/\dot{M}_{\rm{crit}}<1$.
In the limit of $\langle{\dot{M}_{\rm{T}}}\rangle/\dot{M}_{\rm{crit}}\ll1.0$, the dotted 
line corresponds to the solid line of Fig.~\ref{fig:atet}{\it b}.

Which conditions are more promising, $\langle{\dot{M}_{\rm{T}}}\rangle/\dot{M}_{\rm{crit}}<1$ or $\langle{\dot{M}_{\rm{T}}}\rangle/\dot{M}_{\rm{crit}}>1$ ?
The answer is $\langle{\dot{M}_{\rm{T}}}\rangle/\dot{M}_{\rm{crit}}<1$. In order to confirm this answer, 
we consider the minimum value of the critical mass transfer rate in what follows.
{ If $t_{\rm{c}}$ is longer than a Hubble time, the binaries never merge within a Hubble time. 
We, therefore, confine our argument to the case in which $t_{\rm{c}}$ is shorter than a Hubble time.
Since the possible value of $t_{\rm{c}}$ is, then, a Hubble time at a maximum,}
the minimum value of the critical mass transfer rate can be estimated as
\begin{eqnarray}
\dot{M}_{\rm{crit},\rm{min}}
=\frac{q}{(1+q)^2}\frac{m}{l}\left(\frac{m+1}{l}\right)^{-1/3}
\frac{1}{f_{\rm{max}}}\frac{M_{\rm{bh}}}{t_{\rm{H}}},
\label{eq:mcmin}
\end{eqnarray}
where  $t_{\rm{H}}=1/H_{0}\sim1.37\times10^{10}$yr is a Hubble time, $f_{\rm{max}}$ is
the maximum value of $f$ which can be expressed as 
\begin{eqnarray}
f_{\rm{max}}=2\pi\alpha_{\rm{SS}}\left(\frac{m+1}{l}\right)^{-1/3}\left(\frac{c_{\rm{s},1}}{v_{\rm{b},\rm{min}}}\right)^2.
\label{eq:fmax}
\end{eqnarray}
Here $v_{\rm{b},\rm{min}}$ is 
\begin{eqnarray}
v_{\rm{b},\rm{min}}=\sqrt{\frac{GM_{\rm{bh}}}{a_{\rm{h}}}}=\frac{q^{1/2}}{2(1+q)}\sigma,
\label{eq:vmin}
\end{eqnarray}
where $a_{\rm{h}}=G\mu/4\sigma^2$ 
is the hardening radius where the binding energy of the BBH exceeds the kinetic energy of the stars, 
and $\sigma$ is the 1D velocity dispersion of the stars in the galactic nucleus \citep{q96}.
Assuming $\sigma$ approximately equals to the observed steller velocity dispersion, we obtain from
the best current estimate of $M$-$\sigma$ relation \citep{mm05}, 
\begin{eqnarray}
\sigma=\frac{200\rm{kms^{-1}}}{(1.66\pm0.24)^{\beta}}\left(\frac{M_{\rm{bh}}}{10^{8}M_{\odot}}\right)^{\beta}
\label{eq:msigma}
\end{eqnarray}
with $\beta=1/(4.86\pm0.43)$. 
Substituting equations~(\ref{eq:fmax})-(\ref{eq:msigma}) into (\ref{eq:mcmin}), we can obtain
\begin{eqnarray}
\dot{M}_{\rm{crit},\rm{min}}
&=&\frac{2.0\times10^{-1}}{1.37(1.66\pm0.24)^{2\beta}\pi}
\left(\frac{m}{l}\right)\left(\frac{q}{1.0}\right)\left(\frac{0.1}{\alpha_{\rm{SS}}}\right)
\nonumber \\
&\times&
\left(\frac{200\rm{kms^{-1}}}{c_{\rm{s},1}}\right)^{2}\left(\frac{M_{\rm{bh}}}{10^{8}M_{\odot}}\right)^{2\beta+1}\,\left[\frac{M_{\odot}}{\rm{yr}}\right].
\label{eq:mcmin2}
\end{eqnarray}
On the other hand, the Eddington accretion rate $\dot{M}_{\rm{Edd}}$ is 
\begin{eqnarray}
\dot{M}_{\rm{Edd}}=\frac{4\pi{GM_{\rm{bh}}}}{c\kappa_{\rm{T}}}\sim0.2\left(\frac{M_{\rm{bh}}}{10^8M_{\odot}}\right)\,\left[\frac{M_{\odot}}{\rm{yr}}\right],
\label{eq:eddrate}
\end{eqnarray}
where $c$ is the velocity of light and $\kappa_{\rm{T}}$ is the Thompson opacity coefficient.
The ratio of $\dot{M}_{\rm{crit},\rm{min}}$ to $\dot{M}_{\rm{Edd}}$ can be written as
\begin{eqnarray}
\frac{\dot{M}_{\rm{crit,min}}}{\dot{M}_{\rm{Edd}}}
&\sim&
\frac{1}{1.37(1.66\pm0.24)^{2\beta}\pi}\left(\frac{m}{l}\right)\left(\frac{200\rm{kms^{-1}}}{c_{\rm{s},1}}\right)^{2}
\nonumber \\
&\times&
\left(\frac{M_{\rm{bh}}}{10^{8}M_{\odot}}\right)^{2\beta}\left(\frac{0.1}{\alpha_{\rm{SS}}}\right)\left(\frac{q}{1.0}\right).
\end{eqnarray}
Since the sound velocity $c_{\rm{s},1}\sim10\rm{kms^{-1}}$ in the outer region of a typical AGN disk, 
$\dot{M}_{\rm{crit}}^{\rm{min}}$ is always { larger} than $\dot{M}_{\rm{Edd}}$
for supermassive black hole with the mass in the $10^{9}M_{\odot}$ to $10^{6}M_{\odot}$ range.
\citet{haya07} actually confirmed that
the mass transfer rate is sub-Eddington rate if the gas is supplied to the CBD at the Eddington rate.
It is, therefore, promising that most massive BBH evolves towards the orbital decay with the growth of the orbital eccentricity.

\begin{figure*}
\resizebox{\hsize}{!}{
\includegraphics*[width=86mm]{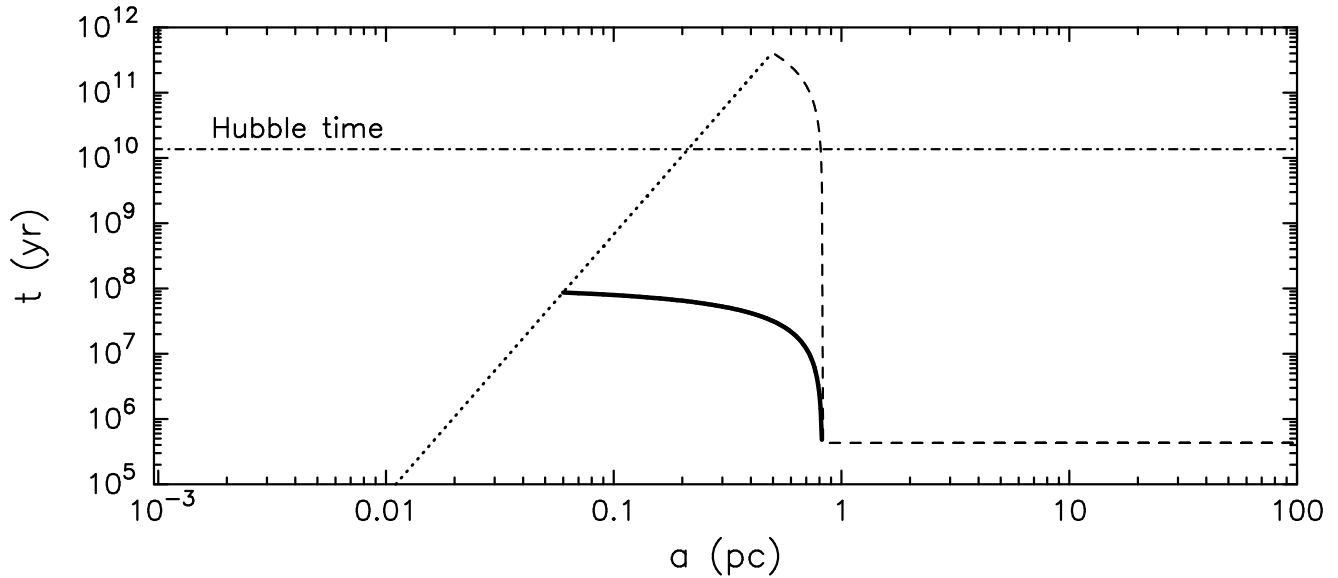}
}
\caption{
Evolutionary timescale of a massive BBH with triple disks 
from $100\rm{pc}$ scale to $10^{-3}\rm{pc}$ scale.
The total black hole mass is $M_{\rm{bh}}=10^{8}M_{\odot}$
with equal mass ratio $M_{2}/M_{1}=1.0$, 
the ratio of the CBD mass to to the total black hole mass is 
$M_{\rm{CBD}}/M_{\rm{bh}}=10^{-2}$, and
the ratio of mass transfer rate to the critical mass transfer rate
is $\langle\dot{M_{\rm{T}}}\rangle/\dot{M}_{\rm{crit}}=0.5$.
The dashed line shows the first evolutionary track 
in which two black holes get close each other 
by their angular momentum loss 
due to the dynamical friction between the black holes 
and the stars surrounding them.
The dotted line shows the third evolutionary track in which 
the BBH finally coalesce 
by the dissipation due to the emission of the gravitational radiation. 
The thick line shows a new, second evolutionary track in which 
the angular momentum of the BBH is removed 
by the binary-disk interaction. 
The horizontal dot-dashed line shows a Hubble time $t_{\rm{H}}\sim1.37\times10^{10}$yr.
}
\label{fig:fpp}
\end{figure*}

\section{Discussion}

We study how a massive BBH with triple disks evolves through the binary-disk interaction.
In the long-term, the BBH resonantly interact with the CBD at the OLR radius via which
the orbital angular momentum is mostly extracted by the CBD.
In the short-term, the gas is transferred from the CBD to each component of the BBH.
This process is repeated every binary orbit during the evolution of the BBH,
by which the accretion disk is formed around each black hole, 
and then each accretion disk viscously evolves.
Some fraction of the angular momentum of each accretion disk 
is finally converted to the orbital angular momentum of the BBH
due to the tidal torque of each black hole.
When the mass transfer rate corresponds to a critical value, 
no BBH evolves because the angular momentum exchange 
caused by these binary-disk interactions is balanced.

{ The direction of BBH evolution is determined} by the critical mass transfer rate.
If the mass transfer rate is smaller than the critical mass-transfer rate, 
the separation between two black holes decays with time whereas 
the orbital eccentricity grows with time and is finally close to 1.0.
The grown-up eccentricity will, then, cause the black hole to plunge into the CBD,
resulting in the enlargement of the gap between the central binary and the CBD (cf. \cite{al94}).
This direct interaction would make the semi-major axis shorten 
because of the friction between the black hole and the gas in the CBD.

What happen if the mass transfer rate is large enough to be over the Eddington accretion rate?
Since the mass transfer rate is, then, larger than the critical mass transfer rate, 
the semi-major axis rapidly increases with time 
whereas the binary orbit become circularized with time.
When the binary separation reaches to the hardening radius, 
the growth of the semi-major axis will be again stalled 
by the dynamical friction with the neighboring stars.
Thus, the BBH is long-lived in this case.

Where is the orbital angular momentum of the BBH finally going?
The angular momentum absorbed by the CBD 
is gradually transferred outward with the gas in the CBD.
The gas around the BBH will, therefore, be dispersed if no gas is supplied to the CBD.
However, if the gas is supplied to the CBD during the BBH evolution, 
the outwardly transferred angular momentum will be removed by a supply source.
This is essentially same problem as how the gas supplies to AGNs.
Again, if the gas is supplied to the CBD at any rate, the CBD is likely to be a quasi-steady state.
Such a state is, however, considered to be different 
from the steady state of accretion disk around a single black hole.
The orbital evolution of the BBH should, therefore, be solved with the evolution of the CBD.
We will tackle this problem by the numerical simulation in the forthcoming paper 
where both the full torque of the BBH and the evolution of the CBD will be taken into account.

Figure~\ref{fig:fpp} shows the coalescent timescale of the massive BBH
with $M_{\rm{bh}}=10^8M_{\odot}$ and $M_{\rm{CBD}}/M_{\rm{bh}}=10^{-2}$.
Two black holes firstly evolve towards the shrinkage of the binary separation 
due to the dynamical friction with the neighboring stars \citep{bege80}.
Next, the orbital evolution is stalled at $a_{\rm{h}}$ 
where the evolutionary timescale by the dynamical friction is 
over a Hubble time even if the emission of the gravitational wave 
dissipates the binding energy of the BBH.
However, the BBH can track a new evolutionary path
due to the binary-disk interaction. 
The new evolutionary path is denoted by the solid thick line of Fig.~\ref{fig:fpp}.
In addition, the rapid growth of the orbital eccentricity allows
the connection to the final evolutionary path where the BBH evolves towards
the coalescence by emitting the gravitational radiation, 
since the timescale of the final evolutionary path
in an eccentric binary is the factor ${ (1-e^2)^{7/2}}$ times shorter than that in a circular binary
\citep{peters64}. The combination of the orbital decay and the eccentricity growth 
makes it possible for the massive BBH to coalesce with a Hubble time.

One of the most interesting problem is how many massive BBHs are there in the Universe.
It is noted from Fig.~\ref{fig:tc} and Fig.~\ref{fig:fpp} that the BBH more rapidly coalesces 
as the black hole mass is less massive and the CBD is more massive.
This suggests that a supermassive BBH with masses more than $\sim10^{8-9}M_{\odot}$ is long-lived as a binary.
Also, the relatively less massive BBH with masses $10^{6-7}M_{\odot}$
would sequentially merge in early Universe and grow to a more massive, single BH.
The host galaxies with such the grown black holes would suffer a major merger, resulting in a supermassive BBH formation. 
If so, we can infer, by using $M$-$\sigma$ relation, 
that such supermassive BBHs are observed in well-developed galaxies with the gas-rich nuclei.

\section{Conclusions}

For the purpose of resolving the final parsec problem,
we study the evolution of a massive BBH with triple disks 
on the parsec/subparsec scale. 
Our main conclusions are summarized as follows:
\begin{enumerate}
\renewcommand{\theenumi}{(\arabic{enumi})}
\item
The orbital evolution of the BBH
is characterized by $t_{\rm{c}}\sim\tau_{\rm{vis},0}^{\rm{CBD}}(M_{\rm{CBD}} / M_{\rm{bh}})^{-1}$.
\item
There is a critical mass-transfer rate, which determines the evolutionary direction of the BBH with triple disks. 
When the mass transfer rate is smaller than the critical one, 
a semi-major axis decays with time whereas the orbital eccentricity grows with time.
The orbital eccentricity is rapidly close to $1.0$
during $t_{\rm{c}}/2$ even if the BBH is initially on a circular orbit, 
while the semi-major axis is 0.0 during $t_{\rm{c}}$ regardless of the orbital eccentricity.
The high value of the orbital eccentricity results in a rapid merger 
within a Hubble time through the emission of the gravitational radiation.
\item
When the mass transfer rate is larger than the critical one, 
a semi-major axis grows with time whereas 
the orbital eccentricity decays with time.
In such a system, the semi-major axis reaches
to the hardening radius of the BBH, typically $a_{\rm{h}}\sim1\rm{pc}$ when $M_{\rm{bh}}=10^{8}M_{\odot}$, 
and its orbit is completely circularized. 
\item
Since the minimum value of the critical mass-transfer rate is larger than 
the Eddington accretion rate of a massive black hole with mass in the $10^{6}M_{\odot}$ to $10^{9}M_{\odot}$ range 
{ as far as the evolutionary timescale is shorter than a Hubble time,}
it is plausible that the mass transfer rate is smaller than the critical one. 
Most BBH, therefore, prefers to evolve towards a rapid 
coalescence within a Hubble time.
\end{enumerate}

\bigskip

We thank anonymous referee for the useful comments and suggestions.
K.H. is grateful to Shin Mineshige, Atsuo~T. Okazaki, 
Shigehiro Nagataki, Keitaro Takahashi, Senndouda Yuiti, 
Norita Kawanaka, Ryouji Kawabata, 
Kohta Murase, Kiki Viderdayanty, and Junichi Aoi 
for helpful discussions. 
The authors thank YITP in Kyoto University, 
where this work was extensively discussed during the YITP-W-05-11 on 
September 20--21, 2005, the YITP-W-06-20 on February 13--15, 2007, 
and the YITP-W-07-20 on January 8--11, 2008. 
The calculations reported here were 
performed using the facility at the Centre for Astrophysics \& Supercomputing 
at Swinburne University of Technology, 
Australia and at YITP in Kyoto University.  This work has been supported in 
part by the Grants-in-Aid of the Ministry of Education, Science, Culture, and 
Sport and Technology (MEXT; 30374218 K.H), 
and by the Grant-in-Aid for the 21st Century COE Scientific Research Programs on 
"Topological Science and Technology'' and "Center for Diversity 
and Universality in Physics'' from MEXT.


\end{document}